\documentstyle[12pt,epsfig]{article}
\begin{document}
\setlength{\baselineskip}{0.65cm}
\begin{titlepage}
\begin{flushright}
RAL-TR-95-059 \linebreak
October 1995
\end{flushright}
\vspace*{2.5cm}
\begin{center}
\Large{
{\bf Radiatively Generated Parton Distributions of} \\
{\bf Polarized Hadrons and Photons*} } \\
\vskip 3.5cm
{\large Werner Vogelsang}  \\
\vspace*{1cm}
\normalsize
Rutherford Appleton Laboratory, \\
Chilton DIDCOT, Oxon OX11 0QX, England
\end{center}
\vskip 2.5cm
\begin{center}
{\bf Abstract} \\
\end{center}
A next-to-leading order QCD analysis of spin asymmetries and structure
functions in polarized deep inelastic lepton nucleon scattering is
presented within the framework of the radiative parton model.
The $Q^2$-dependence of the spin asymmetry $A_1^{p}(x,Q^2)$
is shown to be non-negligible for $x$-values relevant for the analysis
of
the present data. In the second part of the paper we present LO results
for radiatively generated spin-dependent parton distributions of the
photon and for its structure function $g_1^{\gamma}$.
\vskip 2cm
\begin{description}
\item[*] Invited talk presented at the ''Workshop on the Prospects of
Spin Physics at HERA'', \linebreak[4]
\hspace*{-0.75cm} DESY-Zeuthen, Germany, August 28-31, 1995.
\end{description}
\end{titlepage}
\section{Introduction}
The past few years have seen much progress in our knowledge about the
nucleons' spin structure due to the experimental study of the spin
asymmetries
$A_1^N (x,Q^2)\approx g_1^N(x,Q^2)/F_1^N(x,Q^2)$ ($N=p,n,d$) in
deep-inelastic scattering (DIS) with longitudinally polarized lepton
beams
and nucleon targets. Previous data on $A_1^p$ by the SLAC-Yale
collaboration \cite{slac} have been succeeded by more accurate
data from [2-4], which also cover a wider range in $(x,Q^2)$,
and results on $A_1^n$ and $A_1^d$ have been published in \cite{e142}
and \cite{smcd,e143d}, respectively.

On the theoretical side, it has become possible to perform a complete
and consistent study of polarized DIS in
next-to-leading order (NLO) of QCD, since the calculation of the
spin-dependent two-loop anomalous dimensions,
needed for the NLO evolution of polarized parton distributions, has
been completed recently \cite{ref11}. A first such study has been
presented in \cite{grsv}, where the underlying concept has been the
radiative generation of parton distributions from a low resolution
scale $\mu$, which in the unpolarized case had previously led
\cite{ref14} to the remarkably successful prediction of the
small-$x$ rise of the proton structure function $F_2^p$
as observed at HERA \cite{gamma}. The main findings of this NLO analysis
\cite{grsv}, which followed the lines of an earlier leading order (LO)
study \cite{ref1}, will be collected in section 2.

New precise data on polarized DIS will be added
in the near future from the HERMES experiment \cite{herm} at HERA.
Moreover, it is no longer inconceivable to longitudinally polarize
HERA's high-energy proton beam. The corresponding situation with
unpolarized beams has already demonstrated \cite{gamma} that at such
high energies the $ep$ interactions, since dominated by the exchange of
almost real (Weizs\"{a}cker-Williams) photons, can reveal also
information on the parton content of the {\em photon} in addition to
that of the proton. Therefore, a polarized $\vec{e}\vec{p}$-collider
mode
of HERA could in principle serve to explore the spin-dependent parton
distributions of circularly polarized photons. These are completely
unmeasured and thus unknown up to now, and one has to invoke models
\cite{gv}
in order to study their expected size and to estimate the theoretical
uncertainty in predictions for them. For this purpose, it seems
worthwhile to also resort to a radiative generation of the
{\em photon's} polarized parton distributions \cite{gv}, since the
corresponding predictions for
the unpolarized photon \cite{grvg} have again been phenomenologically
successful \cite{gamma}. This topic will be covered in section 3.
\section{NLO Radiative Parton Model Analysis of Polarized DIS}
Measurements of polarized deep inelastic lepton nucleon scattering
yield direct information [1-7] on the spin-asymmetry
\begin{equation}
A_1^N (x,Q^2) \simeq \frac{g_1^N (x,Q^2)}{F_1^N (x,Q^2)}
=\frac{g_1^N(x,Q^2)}{F_2^N (x,Q^2)/ \left[ 2x(1+R^N(x,Q^2))
\right] } \:\:\: ,
\end{equation}
where $N=p,n,d$ and $R \equiv F_L/2xF_1 =(F_2-2 x F_1)/2xF_1$.
In NLO, $g_1^N(x,Q^2)$ is related to the polarized ($\Delta f^N$)
quark and gluon distributions in the following way:
\begin{eqnarray}
\nonumber
g_1^N(x,Q^2) &=& \frac{1}{2} \sum_q e_q^2\; \Bigg\{ \Delta q^N(x,Q^2)+
\Delta \bar{q}^N(x,Q^2)+\\
&+& \frac{\alpha_s(Q^2)}{2\pi} \left[ \Delta C_q * \left( \Delta q^N+
\Delta \bar{q}^N\right) + \Delta C_g * \Delta g\right]
\Bigg\}
\end{eqnarray}
with the convolutions ($*$) being defined as usual,
and where the appropriate spin-dependent Wilson coefficients
$\Delta C_i$ in the
$\overline{\mbox{MS}}$ scheme are given, e.g., in \cite{ref11}.
The NLO form of the unpolarized
(spin-averaged) structure function $F_1^N(x,Q^2)$ is similar to the one
in (2) with $\Delta f^N (x,Q^2)\rightarrow f^N (x,Q^2)$ and
the unpolarized Wilson coefficients given, for example,
in \cite{ref13}.

The NLO $Q^2$-evolution of the spin-dependent parton distributions
$\Delta f(x,Q^2)\equiv \Delta f^p (x,Q^2)$ is performed most
conveniently in
Mellin $n$-moment space where the solutions of the evolution equations
(see, e.g., refs.[16-18]) can be obtained analytically,
once the boundary conditions at some $Q^2=\mu^2$, i.e. the input
densities $\Delta f(x,\mu^2)$ to be discussed below, are specified.
These $Q^2$-evolutions are governed by the spin-dependent
LO \cite{ar} and NLO \cite{ref11} ($\overline{\mbox{MS}}$)
anomalous dimensions.
Having obtained the analytic NLO solutions for the moments of parton
densities it is simple to (numerically) Mellin-invert
them to Bj\o rken-$x$ space as described, for example, in
\cite{ref16,ref17}.
As seen in (2), the so obtained $\Delta f(x,Q^2)$ are then convoluted
with the Wilson coefficients $\Delta C_i$ to yield the desired
$g_1(x,Q^2)$.

To fix the polarized NLO input parton distributions
$\Delta f(x,Q^2=\mu^2)$ we perform fits to the directly measured
asymmetry
$A_1^N(x,Q^2)$ in (1), rather than to the derived $g_1^N(x,Q^2)$,
mainly because for the experimental extraction of the latter often the
assumption of the $Q^2$-independence of $A_1^N(x,Q^2)$ is made, which is
theoretically not justified as we will see below.
As mentioned in the introduction, the other main ingredient of our NLO
analysis \cite{grsv} is that we follow the radiative (dynamical) concept
\cite{ref14} by choosing the low input scale $Q^2=\mu^2=0.34\,
{\mbox{GeV}}^2$
and implementing the fundamental positivity constraints
\begin{equation}
|\Delta f(x,Q^2)| \leq f(x,Q^2)
\end{equation}
down to $Q^2=\mu^2$. Therefore we shall use all presently available data
[2-7] in the small-$x$ region where
$Q^2\,$\raisebox{-1mm}{${\stackrel{\textstyle >}{\sim}}$}$\,1\,
{\mbox{GeV}}^2$, without introducing lower cuts in $Q^2$ as was usually
necessary in previous analyses \cite{ref28}.
A further advantage of this approach is the possibility to study the
$Q^2$-dependence of $A_1^N(x,Q^2)$ over a wide range of $Q^2$ which might
be also relevant for possibly forthcoming polarization experiments
at HERA.
The analysis affords some well established set of unpolarized NLO parton
distributions $f(x,Q^2)$ for calculating $F_1^N(x,Q^2)$ in (1) which will
be adopted from ref.\cite{ref14}.

In addition to (3), the polarized NLO parton distributions
$\Delta f(x,Q^2)$
are, for the $SU(3)_f$ symmetric 'standard' scenario, constrained by
the sum rules
\begin{eqnarray}
&&\int_0^1 dx \left( \Delta u + \Delta \bar{u} -
\Delta d - \Delta \bar{d}
\right) (x,\mu^2) = g_A = F + D = 1.2573 \pm 0.0028 \\
&&\int_0^1 dx \left( \Delta u + \Delta \bar{u} +
\Delta d + \Delta \bar{d} -
2(\Delta s + \Delta \bar{s}) \right) (x,\mu^2) = 3F -D = 0.579 \pm 0.025
\end{eqnarray}
with the values of $g_A$ and $3F-D$ taken from \cite{ref25}.
It should be noted that the first moments of the flavor
non-singlet combinations which appear in (4) and (5) are
$Q^2$-independent also in NLO \cite{kod,grsv,alex}.

As a plausible alternative to the full $SU(3)_f$ symmetry between charged
weak and neutral axial currents required for deriving the 'standard'
constraints (4) and (5), we consider a 'valence' scenario
\cite{ref1,ref2}
where this flavor symmetry is broken and which is based on the assumption
\cite{ref2} that the flavor changing
hyperon $\beta$-decay data fix only the total helicity of
{\em{valence}} quarks:
\setcounter{equation}{3}
\renewcommand{\theequation}{\arabic{equation}'}
\begin{eqnarray}
\int_0^1 dx \left( \Delta u_v -
\Delta d_v \right) (x,\mu^2) &=& g_A = F+D= 1.2573 \pm 0.0028 \\
\int_0^1 dx \left( \Delta u_v +\Delta d_v \right) (x,\mu^2)
&=& 3F-D  = 0.579 \pm 0.025\;\;\;.
\end{eqnarray}
We note that in both above scenarios the Bj\o rken sum rule
manifestly holds due to the constraints (4),(4'). Our optimal NLO input
distributions at $Q^2=\mu^2=0.34\,{\mbox{GeV}}^2$ subject to these
constraints can be found in \cite{grsv}.

A comparison of our results with the data on $A_1^N(x,Q^2)$ is
presented in
Fig.1. Obviously, very similar results are obtained for the two scenarios
considered above.
The $Q^2$-dependence of $A_1^p (x,Q^2)$ is presented in Fig.2 for
some typical fixed $x$ values for $1\le Q^2\le 20 {\mbox{GeV}}^2$
relevant
for present experiments. In the $(x,Q^2)$ region of present data
[2-7], $A_1^p(x,Q^2)$ increases with
$Q^2$ for $x>0.01$. Therefore, since most present data in the small-$x$
region correspond to small values of $Q^2\,
$\raisebox{-1mm}{${\stackrel{\textstyle >}{\sim}}$}$\,1$\,GeV$^2$,
the determination of $g_1^p(x,Q^2)$ at a larger fixed
$Q^2$ (5 or 10 GeV$^2$, say) by assuming $A_1^p(x,Q^2)$ to be independent
of $Q^2$, as is commonly done [2-7], is misleading and might lead to
an {\em{under}}estimate of $g_1^p$ by as much as about $20\%$. Results
for the (also non-negligible) $Q^2$-dependence of $A_1^n$ can be found
in \cite{grsv,ref1}. The assumption of approximate scaling for
$A_1 (x,Q^2)$ is therefore unwarranted and, in any case, theoretically
not justified \cite{grsv,ref1,ref27,ref29}.

\begin{figure}[p]
\vspace*{-1cm}
\hspace*{-1.5cm}
\epsfig{file=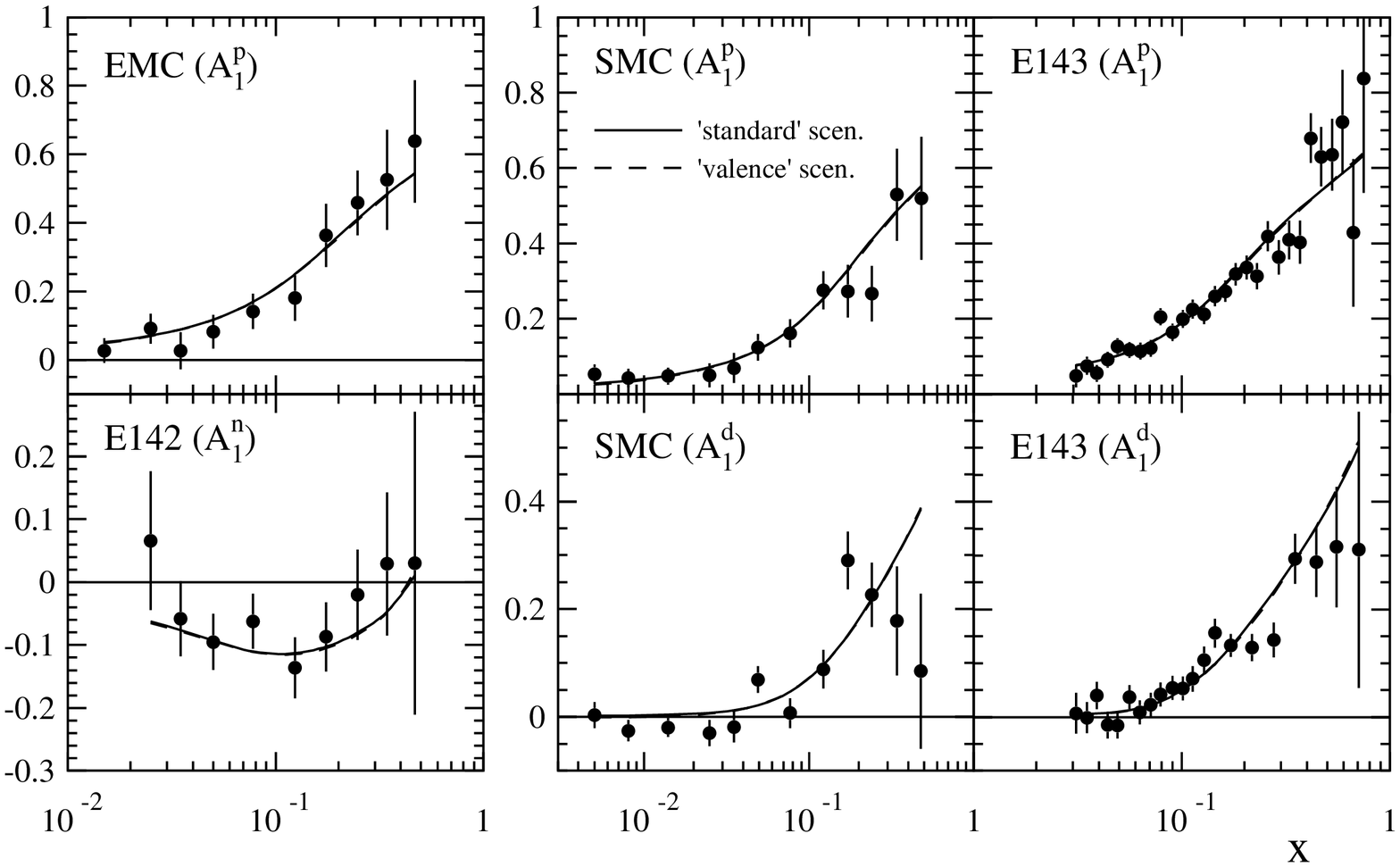,width=19.7cm,angle=0}
\vspace*{-1cm}
\caption{ {\sf Comparison of our NLO results for \protect{$A_1^{N}
(x,Q^2)$}
as obtained from the fitted inputs at \protect{$Q^2=\mu^2$}
for the 'standard' and 'valence' scenarios with present data [2-7].
The \protect{$Q^2$} values adopted here correspond to the different
values quoted in [2-7] for each data point.}}
\vspace*{-0.2cm}
\hspace*{0.7cm}
\epsfig{file=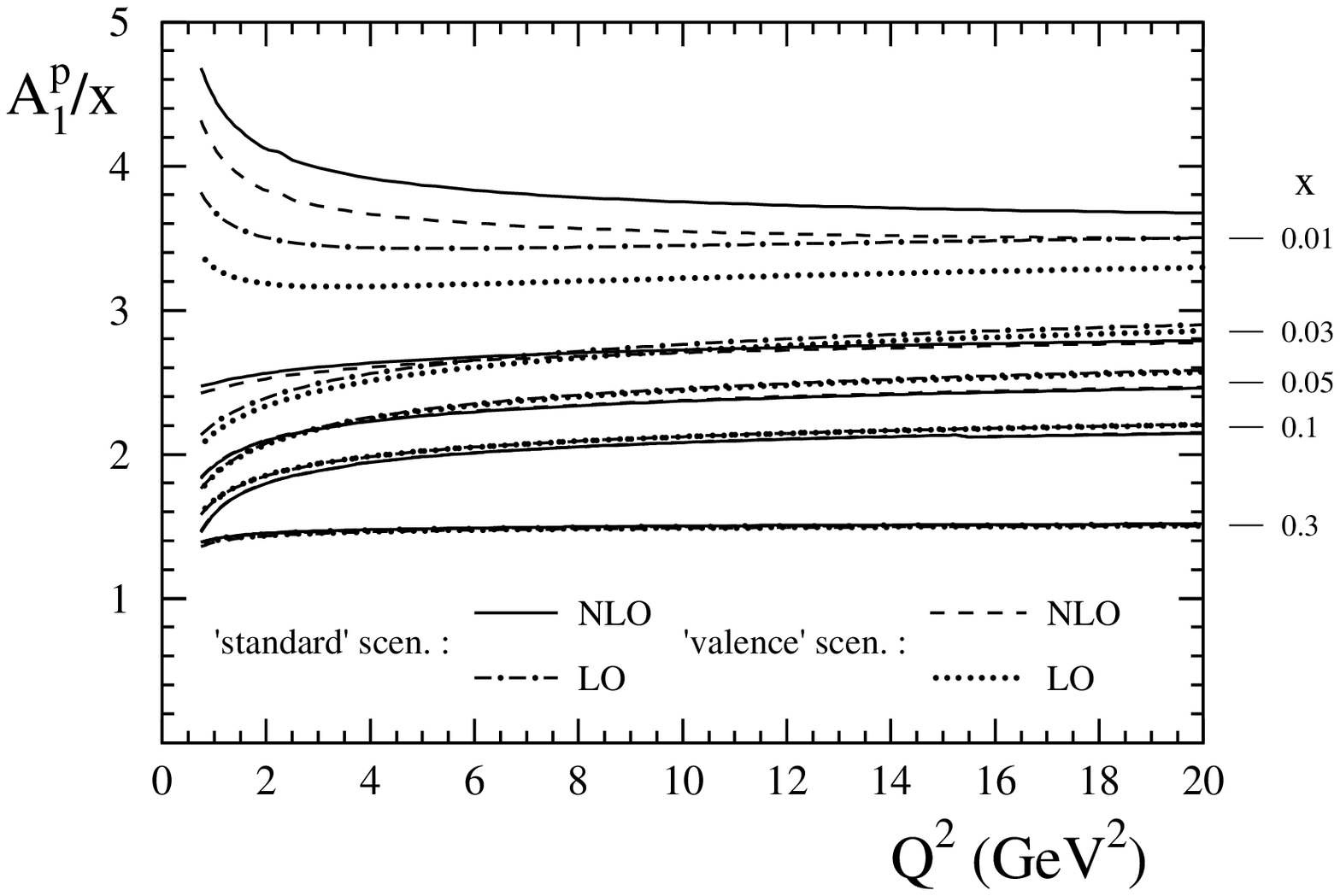,width=14.7cm,angle=0}
\vspace*{-1.9cm}
\caption{ {\sf The \protect{$Q^2$}-dependence of
\protect{$A_1^p (x,Q^2)$}
as predicted by the NLO QCD evolution at various fixed values of
\protect{$x$}. The LO results are from [12].}}
\vspace*{-0.3cm}
\end{figure}
A further result of our analysis is that the polarized gluon density
$\Delta g(x,Q^2)$ is hardly constrained by present experiments.
Similarly agreeable fits as those shown in Fig.1 to all present
asymmetry data can also be obtained for a fully saturated gluon input
$\Delta g(x,\mu^2)=g(x,\mu^2)$ as well as for the less saturated
$\Delta g(x,\mu^2)=xg(x,\mu^2)$ or even a purely dynamical
input $\Delta g(x,\mu^2)=0$.
We compare such gluons at $Q^2=4$ GeV$^2$ in Fig.3. The variation
of $\Delta g(x,Q^2)$ allowed by present experiments is indeed
sizeable. This implies, in particular, that the $Q^2$-evolution of
$g_1(x,Q^2)$ below the experimentally accessible $x$-range is not
predictable for the time being.

To conclude this section, our results \cite{grsv,ref1} demonstrate
the compatibility of the restrictive radiative model \cite{ref14}
with present measurements of deep inelastic spin asymmetries and
structure functions.
\vspace*{-1.2cm}
\begin{figure}[h]
\hspace*{1.3cm}
\epsfig{file=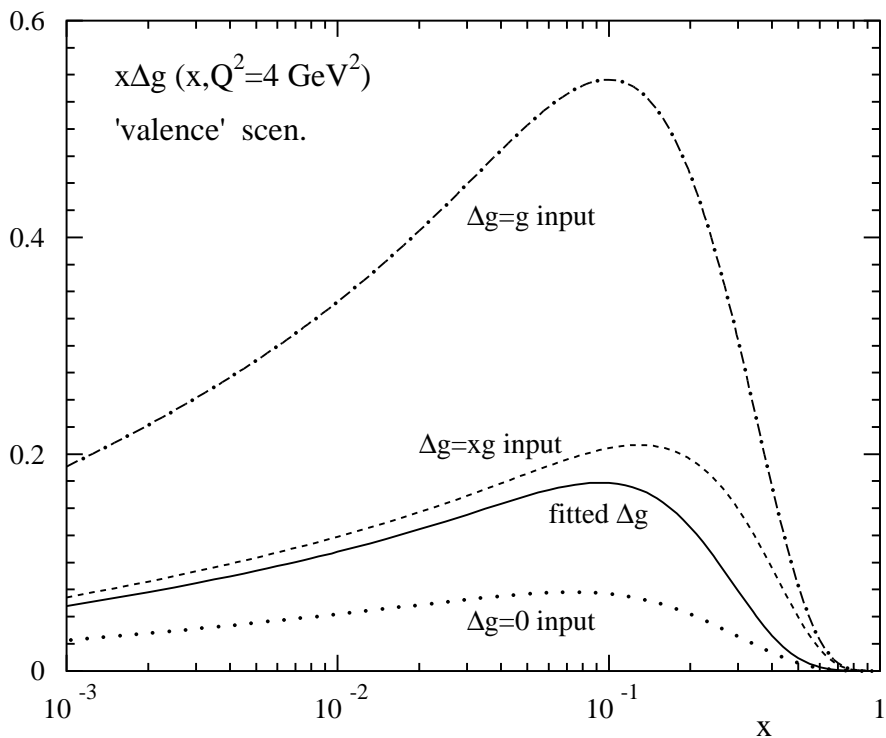,width=13.7cm,angle=0}
\vspace*{-1.3cm}
\caption{ {\sf The experimentally allowed range of polarized gluon
densities at \protect{$Q^2=4\,{\mbox{GeV}}^2$} for the 'valence'
scenario
with differently chosen \protect{$\Delta g(x,\mu^2)$} inputs.}}
\vspace*{-0.3cm}
\end{figure}
\section{Spin-dependent parton distributions of the photon}
As mentioned in the introduction, the photon's polarized parton
distributions,
$\Delta f^{\gamma}=\Delta q^{\gamma}$, $\Delta g^{\gamma}$,
are presently unmeasured and thus completely unknown. Theoretical
expectations can, however, be derived \cite{gv} by assuming a
radiative generation of the $\Delta f^{\gamma}$
along the lines followed \cite{grvg} successfully for the unpolarized
photonic parton distributions, $f^{\gamma}=q^{\gamma},g^{\gamma}$. In
\cite{grvg} a VMD valence-like structure at the low resolution
scale $Q^{2}=\mu^{2}=0.25$ GeV$^2$ was imposed as the input boundary
condition, assuming that at this resolution scale the photon behaves
like a
vector meson, i.e., that its parton content is proportional to that
of the $\rho$-meson. Since nothing is known experimentally about
the latter, the parton densities of the neutral pion as determined in
a previous study \cite{grvp} were used instead which are expected not
to be too dissimilar from those of the $\rho$. Unfortunately, this
procedure
is obviously impossible for determining the VMD input distributions
$\Delta f^{\gamma}(x,\mu^2)$ for the {\em polarized} photon. Some help
is again provided by the positivity constraints
\setcounter{equation}{5}
\renewcommand{\theequation}{\arabic{equation}}
\begin{equation}
|\Delta f^{\gamma}(x,\mu^{2})|\leq f^{\gamma}(x,\mu^{2})  \:\:\: ,
\end{equation}
and it is interesting to see how restrictive these general conditions
already are. For this purpose we consider \cite{gv}
two very different scenarios
with 'maximal', $\Delta f^{\gamma}(x,\mu^2)=f^{\gamma}(x,\mu^2)$,
and 'minimal', $\Delta f^{\gamma}(x,\mu^2)=0$, saturation of (6).
We mention that a sum rule expressing the vanishing of the
first moment of the polarized photon structure function
$g_1^{\gamma}$ was derived from current conservation in \cite{sr} which,
in the LO considered here, is equivalent to
\begin{equation}
\int_0^1 \Delta q^{\gamma}(x,\mu^2)=0 \:\:\: .
\end{equation}
This sum rule can in principle serve to further restrict the range of
allowed VMD inputs. On the other hand, we are interested only in the
region of, say, $x>0.01$ here, such that the current conservation
constraints (7) could be implemented by contributions from smaller $x$.
To estimate the
uncertainties in the predictions for the $\Delta f^{\gamma}(x,Q^2)$
stemming from the insufficiently known VMD input we therefore stick to
the two extreme scenarios discussed above, even though strictly speaking
the maximally saturated input violates the sum rule (7).

\begin{figure}[p]
\vspace*{-1cm}
\hspace*{-1cm}
\epsfig{file=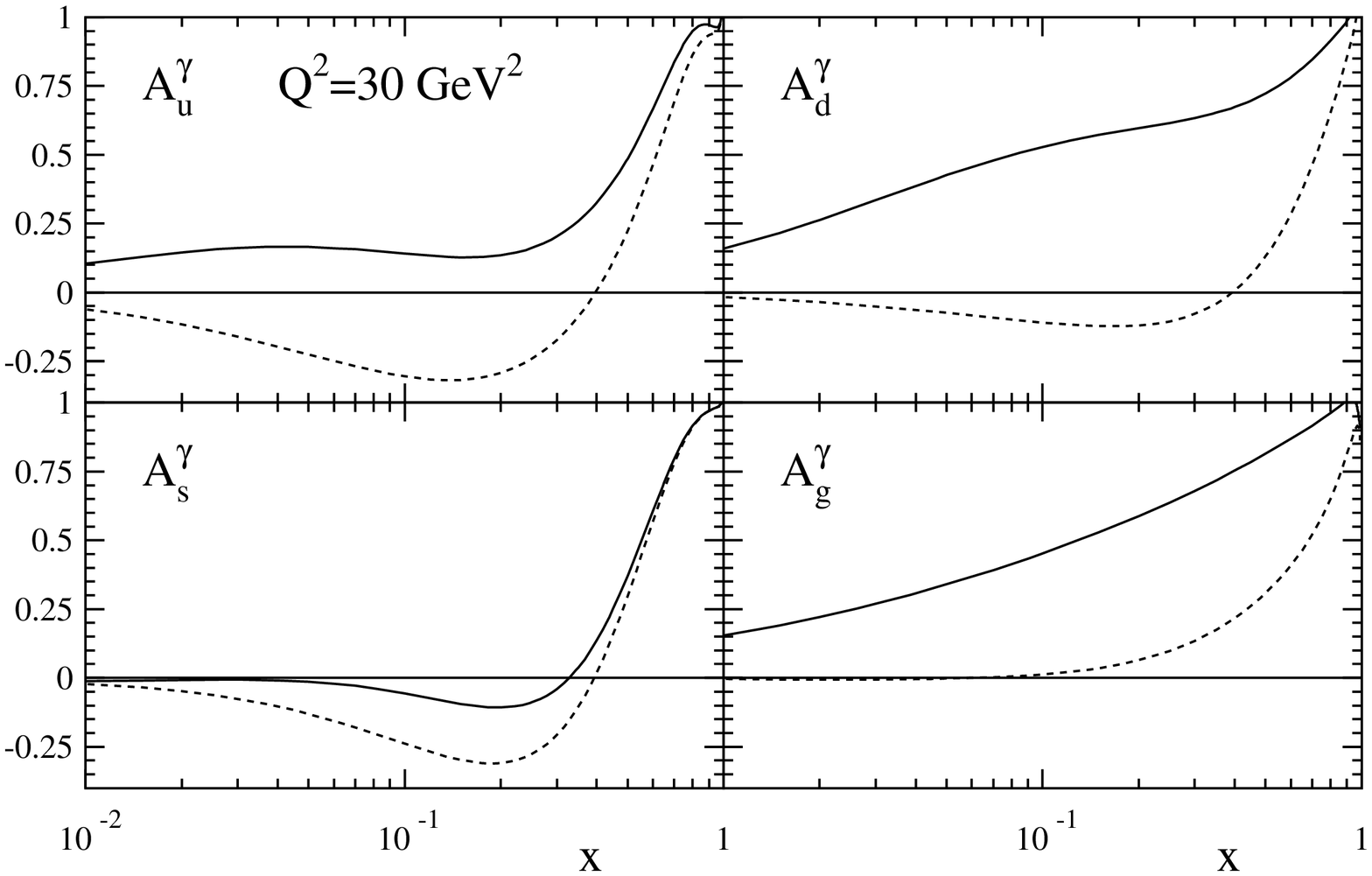,width=16.7cm,angle=0}
\vspace*{-0.6cm}
\caption{ {\sf The predicted photonic parton asymmetries
\protect{$A_f^{\gamma} (x,Q^2)$} at \protect{$Q^2=30$ GeV$^2$}
as defined in (8) for 'maximal' (solid lines) and 'minimal'
(dashed lines)
saturation of the VMD input.}}
\vspace*{-0.5cm}
\hspace*{2.2cm}
\epsfig{file=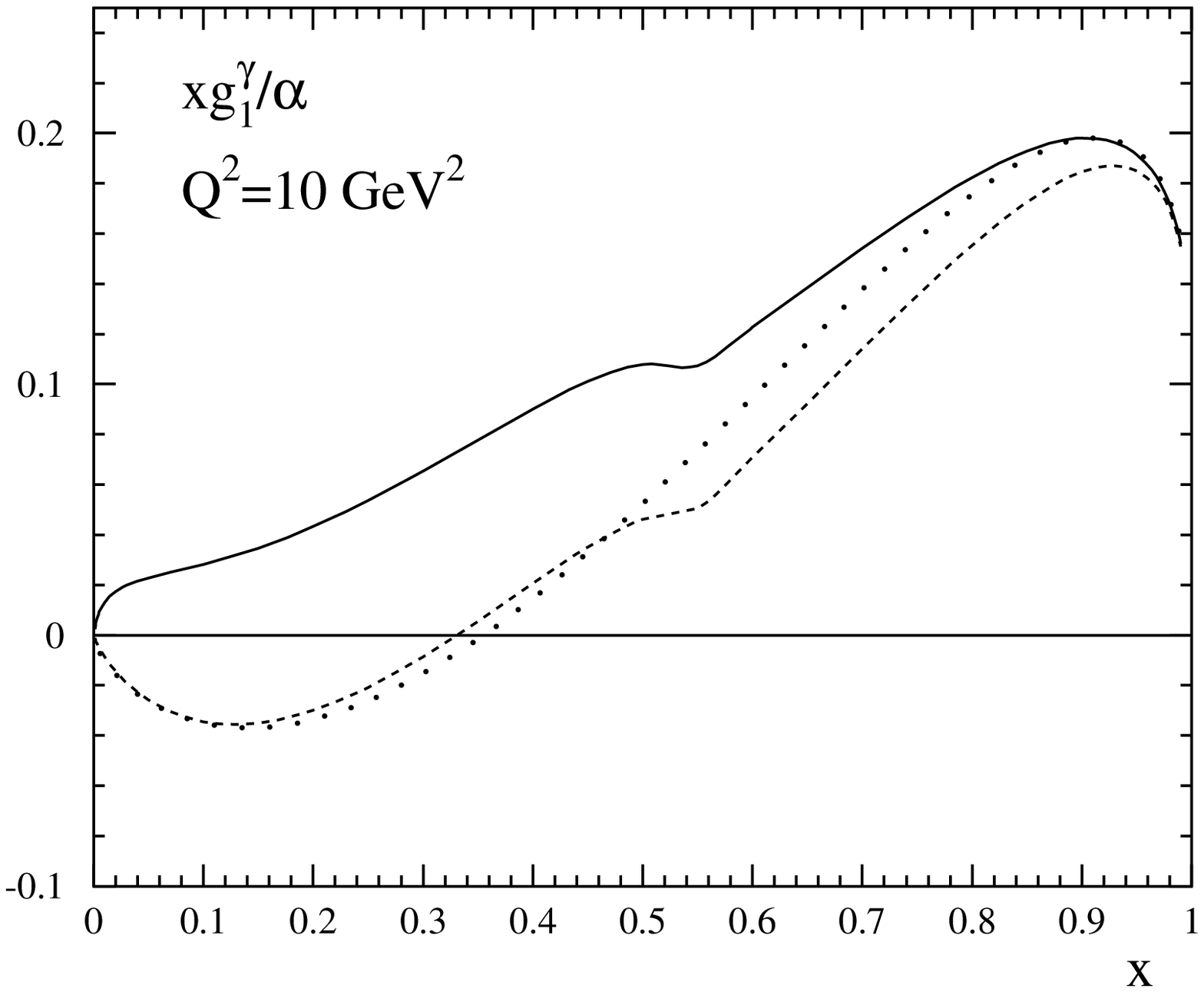,width=12.7cm,angle=0}
\vspace*{-1cm}
\caption{ {\sf LO predictions for the photon's spin-dependent structure
function \protect{$xg_1^{\gamma}(x,Q^2)/\alpha$} at
\protect{$Q^2=10$ GeV$^2$} for 'maximal' (solid) and 'minimal' (dashed)
saturation of the VMD input. The dotted line shows the result for
the LO 'asymptotic' solution for 3 active flavors also considered
in [31].}}
\vspace*{-0.3cm}
\end{figure}
Starting from the two different boundary conditions for
$\Delta f^{\gamma}(x,\mu^{2})$ we generate $\Delta f^{\gamma}(x,Q^{2})$
at $Q^{2}>\mu^{2}$ as in \cite{grvg} replacing the unpolarized splitting
functions in the $Q^{2}$-evolution equations by their polarized
counterparts \cite{ar}. In view of the uncertainties in the input, we
restrict our calculations to the leading order, although in principle
a NLO analysis has become possible now \cite{fut} by exploiting the
results for the polarized two-loop splitting functions of ref.
\cite{ref11}.
The resulting parton asymmetries \cite{gv}
\begin{equation}
A_{f}^{\gamma}(x,Q^{2})\equiv \frac{\Delta f^{\gamma}(x,Q^{2})}
{f^{\gamma}(x,Q^{2})}
\end{equation}
are shown in Fig.4 for $Q^2=30$ GeV$^2$. As can be seen, there are
quite substantial differences between the results for the two
scenarios. Fig.5 shows the corresponding LO results for the polarized
photon structure function
\begin{equation}
g_1^{\gamma}(x,Q^2) \equiv \sum_{q} e_q^2 \Delta q^{\gamma}
(x,Q^2) \:\:\: ,
\end{equation}
where charm contributions from the subprocesses $\gamma^{\ast}\gamma
\rightarrow c\bar{c}$ and $\gamma^{\ast}g\rightarrow c\bar{c}$ have been
included (see \cite{gsv} for further details). For comparison we also
include in Fig.5 the 'asymptotic' result for $g_1^{\gamma}$ which was
also considered in refs. \cite{asy}.
$g_1^{\gamma}$ would in principle be accessible in polarized $e^+e^-$
collisions \cite{gsv} or in $\vec{e}\vec{\gamma}$ processes.
In a polarized
collider mode of HERA, the spin-dependent parton distributions of the
photon will show up in $\vec{\gamma}\vec{p}$ processes like
jet \cite{kun} or heavy flavor photoproduction when the photon is
resolved into its hadronic structure \cite{fut}.
\section*{Acknowledgements}
I am thankful to M. Gl\"{u}ck, E. Reya, and M. Stratmann for a fruitful
collaboration.

\end{document}